\documentclass{osa-article}
\journal{boe}

\newcommand{\mymarginpar}[1]{} 

\renewcommand{\vec}[1]{\mathbf{#1}}

\RequirePackage{bm}
\def\mat#1{\ensuremath{\bm#1}}

\newcommand{\ts}{}

\begin{document}

\title{Efficient and high accuracy 3-D OCT angiography motion correction in pathology}

\author{Stefan~B.~Ploner,\authormark{1,2,*}
Martin~F.~Kraus,\authormark{1}
Eric~M.~Moult,\authormark{2}
Lennart~Husvogt,\authormark{1,2}
Julia~Schottenhamml,\authormark{1}
A.~Yasin~Alibhai,\authormark{3}
Nadia~K.~Waheed,\authormark{3}
Jay~S.~Duker,\authormark{3}
James~G.~Fujimoto,\authormark{2}
and Andreas~K.~Maier\authormark{1}}

\address{\authormark{1}Pattern Recognition Lab, Friedrich-Alexander-Universit\"at Erlangen-N\"urnberg, Martensstr. 3, Erlangen, 91058, Germany\\
\authormark{2}Department of Electrical Engineering and Computer Science and Research Laboratory of Electronics, Massachusetts Institute of Technology, 77 Massachusetts Ave, Cambridge, MA 02139, USA\\
\authormark{3}New England Eye Center, Tufts Medical Center, 800 Washington St. Box 450, Boston, MA 02111, USA}

\email{\authormark{*}stefan.ploner@fau.de}

\begin{abstract}
We propose a novel method for non-rigid 3-D
motion correction of orthogonally raster-scanned optical coherence tomography angiography volumes. This is the first approach that aligns predominantly axial structural features like retinal layers and transverse angiographic vascular features in a joint optimization. Combined with the use of orthogonal scans and favorization of kinematically more plausible displacements, the approach allows subpixel alignment
and micrometer-scale distortion correction in all
3
dimensions.
As no specific structures
or layers
are segmented, the approach is by design robust to pathologic changes.
It is furthermore designed for highly parallel implementation and brief runtime, allowing its integration in clinical routine even for
high density or
wide-field scans.
We evaluated the algorithm with metrics related to clinically relevant features in a large-scale quantitative evaluation based on 204 volumetric scans
of 17 subjects
including both a wide range of pathologies and healthy controls.
Using this method, we
achieve state-of-the-art axial performance and show significant advances in both transverse co-alignment and distortion correction, especially in the pathologic subgroup.
\end{abstract}

\section{Introduction}

\mymarginpar{motivation}
Optical coherence tomography (OCT) is a non-invasive 3-D optical imaging modality demonstrated and widely used in ophthalmology\cite{Huang1991, Fujimoto2016}. Since the introduction of Fourier-domain OCT\cite{Wojtkowski2002}, 1-D depth profiles are successively acquired over a surface to form a 3-D volume. Typically, the surface is scanned line by line, where each line scan forms a cross-sectional image or a B-scan. Since B-scans are acquired in milliseconds, slices extracted along a scan line, or the fast scan axis, are barely affected by motion. In contrast, slices extracted orthogonally to scan lines, i.\,e.\ in slow scan direction, are affected by various types of eye motion occurring throughout the full, multi-second acquisition time.
Since they both persist during fixation, the most relevant types of eye movements during scanning are
(micro-)saccades, which can introduce discontinuities or gaps between B-scans, and slow drifts, which cause small, slowly changing distortion\cite{MartinezConde2004}.
Further eye motion is caused by pulsatile blood flow, respiration and head motion.
Despite ongoing advances in scanning speed\cite{Potsaid2010, Klein2017}
typical volume acquisition times remain at a constant level.
Instead, the additional scanning speed is used for dense volumetric scanning or wider fields of view\cite{Kolb2015}. Especially OCT angiography (OCTA)\cite{Makita2006, Wang2007, Mariampillai2008, Jia2012} multiplies the required number of samples by at least two, a number which may increase further to accommodate recent developments in blood flow speed estimation which are based on multiple interscan times\cite{Tokayer2013, Ploner2016}. As a consequence, an ongoing need for improvement in motion compensation especially in pathology is widely agreed on in literature\cite{SanchezBrea2019, Enders2019, Lauermann2018}.

\mymarginpar{state of the art}
One approach for correcting such motion is the use of additional hardware, e.g. a scanning laser ophthalmoscope, for tracking and correction of the scan position\cite{Ferguson2004, Vienola2012, LaRocca2013} or retrospective alignment with a reference image\cite{Ricco2009}. While only tracking provides the capability of a gap-free acquisition with a single scan, disadvantages are e.\,g.\ increased system complexity and cost,
increased scan duration if scanning is interrupted due to saccadic eye motion,
and limited accuracy, which is especially important in the context of OCT angiography. The second approach is to acquire multiple OCT scans and to perform a software-based alignment retrospectively. While this increases scan time and delays the display of the corrected result, accuracy down to below OCT resolution is possible and individual scans can be merged to create higher quality images\cite{Kraus2012, Heisler2017}. Some methods additionally allow mosaicing of partially overlapping volumes\cite{Hendargo2013, Zang2016}, but are limited to alignment in the en face plane. Others allow full 3-D alignment including the axial dimension\cite{Kraus2012, Kraus2014, Lezama2016, Zang2017, Chen2017}. A fundamental differentiation between software methods is the underlying scan pattern. Methods that use multiple raster scans with their B-scans oriented along the same direction easily allow the registration of non-square field-of-views\cite{Zang2016, Zang2017, Heisler2017}. However, since there is no motion-free reference in the direction orthogonal to the B-scans, they are limited to remove uncorrelated saccadic motion.
Spatially correlated saccadic motion may occur when patients involuntarily fixate the scanning beam while it traverses the fovea.
Drift motion is only aligned with a reference but not corrected, propagating thereby caused distortion from the reference to the result. In contrast, methods based on orthogonally oriented B-scans\cite{Kraus2012, Hendargo2013, Kraus2014, Lezama2016} or lissajous scans\cite{Chen2017, Chen2018} have almost motion-free slices available along all dimensions and thus have the potential for most accurate motion correction. Finally, it is common to do at least simple software-based registration before the OCT angiography computation irrespective of the use of tracking, and it was reported that the combination of both approaches can be superior to hardware or software correction alone\cite{Camino2016}.

\mymarginpar{why we extend kraus}
This paper extends the orthogonal raster scan based method by Kraus et~al.\ \cite{Kraus2014}.
Distinct features of this method are the joint motion correction of all dimensions in a combined optimization and a displacement field model that is designed along the OCT scanning process: First, instead of having a single displacement field that maps a \textit{moving} volume to a \textit{reference}, one displacement field exists for each volume. This enables the method to correct the motion in all scans, potentially avoiding propagation of drift motion to the final result. Secondly, the displacement fields are enforced to be smooth not isotropically, as it is common for standard registration algorithms, but instead along the scan pattern underlying each individual scan.
This favors estimation of more realistic displacement fields. Combined, these design decisions enable and effect the separation of the displacement into the motion underlying each raster scan, allowing to undistort each individual scan in contrast to a mere co-alignment that may reproduce distortion from the input.

\mymarginpar{proposed advances}
While these features retain their potential for state-of-the-art software motion correction in daily clinical practice to date, since publication in 2014, requirements for motion correction increased with the demand for denser scans and OCTA. In contrast to competing methods, angiography data was not yet incorporated into the registration, and its specifics, like omitting areas with saturated signal due to saccadic motion, were not covered. The fine structure of vessel networks, as visualized by OCTA, at the same time requires but also provides potential for further improved registration accuracy. Additionally, longer scan times for OCTA and higher densities increase distortion from drift and saccadic motion in the uncorrected scans. It is thus the goal of this paper to extend the previous method to utilize the potential of OCTA with minimal impact on runtime, and to demonstrate further improvements in distortion correction capability of the resulting method on dense OCTA scans in a comprehensive and clinically relevant study cohort.

\section{Methods} \label{method}

\mymarginpar{method overview}
This section proposes an extension to the motion correction approach of Kraus et~al.\ \cite{Kraus2012, Kraus2014} to additionally employ OCTA data during registration. In contrast to the OCT volume, the OCTA data is projected to an en face image before it is used in the registration. This is because the layered structure of the eye already provides ample features for axial subpixel registration with standard OCT alone. Due to the retinal curvature especially around the foveal avascular zone and optic nerve, these features are also useful for a global transverse registration. However, these features are continuously changing in the en face plane, limiting therewith achievable accuracy. In contrast, OCTA can provide local features by its high contrast visualization of fine vessel structures. Thus, while the en face approach maintains the relevant benefits of the OCTA signal, the influence on runtime for handling the 2-D OCTA data during optimization is negligible compared to the 3-D OCT data, provided that the en face image can be computed quickly.

\begin{figure}[!ht]
	\centering	\includegraphics{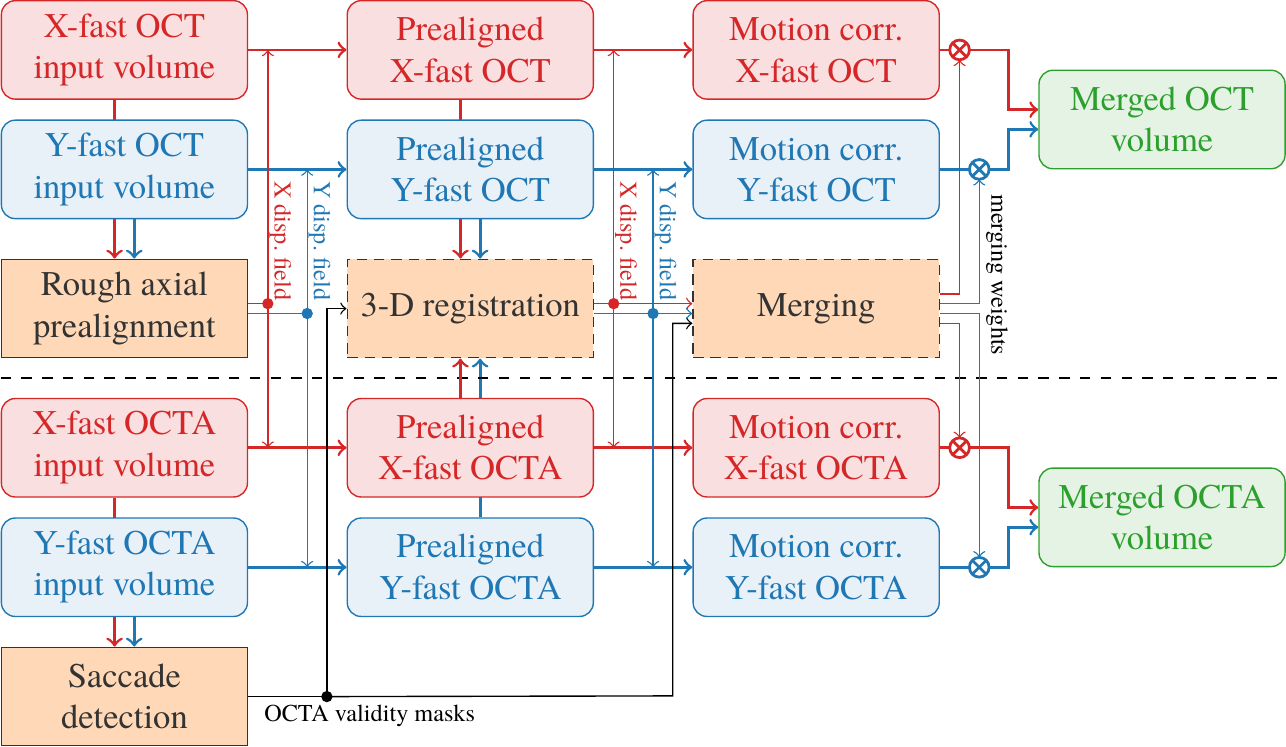}		\caption{Algorithm data flow. The part above the dashed line matches the existing \textit{OCT only} method: First, a rough prealignment and tilt correction is performed in the axial dimension. Then, the volumes are registered jointly in all dimensions and subsequently merged by a weighted average. The proposed \textit{OCT + OCTA} approach adds saccade detection, the 3-D registration is extended to project and use OCTA data, and the merging step is adapted to exclude white line artifacts in OCTA data.}	\label{fig:algorithm-overview}\end{figure}

\mymarginpar{section overview}
\figurename~\ref{fig:algorithm-overview} outlines the algorithm as a whole in terminology used throughout this paper. The \textit{OCT only} algorithm of Kraus et~al.\ \cite{Kraus2012, Kraus2014}
is briefly described in section~\ref{oct-only-algorithm} and introduces basic notations. It is followed by detailed descriptions of the \textit{OCT + OCTA} extension, starting with the detection of motion artifacts in the OCTA data in~\ref{saccade_handling}, continuing with the efficient generation of en face OCTA images in~\ref{OCTA-projection} and their incorporation into the alignment optimization in~\ref{OCTA-similarity}. Finally, merging of the registered OCTA volumes into a single result volume is described in~\ref{octa-merging}.

\subsection{Basic OCT only algorithm} \label{oct-only-algorithm}

\mymarginpar{input}
The basic method uses at least two raster scans of an arbitrary OCT scanner, of which at least one volume must be scanned with B-scans oriented orthogonally to other volumes, such that scans with little motion are available in both lateral dimensions. For simplicity, this paper is limited to the case of exactly two orthogonal scans unless otherwise stated. For a straightforward generalization to more than two volumes see section 2.4 in \cite{Kraus2012}.

\mymarginpar{definitions}
The motion corrected volume shall resemble a hypothetical scan that was performed without motion. It is scanned over
an equiangular
grid with width $w$ and height $h$ with A-scans of depth $d$ such that the voxel positions are located at $\vec p_{ijk} = (x_i, y_j, z_k)$ with $i \in \{ 1, \hdots, w \}$, $j \in \{ 1, \hdots, h \}$, $k \in \{ 1, \hdots, d \}$. According to typical viewing of OCT data, voxel spacing within each axis is assumed to be uniform.
Since the actual acquisitions are distorted by motion, they need to be resampled at these locations to become aligned with the motion-free scan. Thus, the goal of the registration is to estimate the displacements $\mat D$ composed of vectors $\vec d_{ij}^V = \big( \Delta x_{ij}^V, \Delta y_{ij}^V, \Delta z_{ij}^V \big)$,
where $(i,j)$ are the horizontal and vertical A-scan indices and $V \! \in \! \{ X, Y \}$ identifies the X- or Y-fast scan.
It is assumed that there is no distortion within A-scans.

\mymarginpar{objective function \& data term}
First, the registration steps \textit{rough axial prealignment} and \textit{3-D registration} (see \figurename~\ref{fig:algorithm-overview}) are described, which follow the same concept. The estimated displacements are iteratively refined by minimizing an objective function
\begin{align} \label{objective-function}
	\text{arg\,min}_{\mat D} \; S_0^\text{str}(\mat S^X, \mat S^Y, \mat D) + \alpha R(\mat D) + M(\mat D) + T(\mat S^X, \mat S^Y, \mat D),
\end{align}
composed of a data term $S$ that describes the dissimilarity of the motion corrected scans, a regularizer $R$ that penalizes fast displacement changes, the mean shift regularizer $M$ and the tilt normalization term $T$. The constituents are further described in the following.

Alignment is quantified based on the residual difference
\begin{align} \label{eq:residual}
	R_{ijk}(\mat S^X, \mat S^Y, \mat D) =~
		& I\left(\mat S^X, x_i + \Delta x_{ij}^X, y_j + \Delta y_{ij}^X, z_k + \Delta z_{ij}^X)\right) - \nonumber \\
		& I\left(\mat S^Y, x_i + \Delta x_{ij}^Y, y_j + \Delta y_{ij}^Y, z_k + \Delta z_{ij}^Y\right)
\end{align}
of each resampled voxel $i,j,k$, where $I(\mat S^V, x, y, z)$ is an interpolator that samples $\mat S^V$ at position $(x, y, z)$ via Hermitian interpolation and $\mat S^X, \mat S^Y$ are the preprocessed structural OCT data of the X- and Y-fast input volumes. The purpose of the preprocessing is to normalize the data to become invariant to noise and illumination changes from e.g.\ vitreous opacities.
After applying a loss function $L$, the sum over all voxels is computed to form the structural similarity\begin{align} \label{oct-only-similarity}
	S_0^\text{str}(\mat S^X, \mat S^Y, \mat D) &= \sum_i \sum_j \sum_k L\left(R_{ijk}(\mat S^X, \mat S^Y, \mat D)\right).
\end{align}
$L$ was chosen to be the slightly modified pseudo-huber-loss
\begin{align} \label{pseudo-huber-norm}
	L_{H,\epsilon_H}(x) = \epsilon_H \cdot \left( \sqrt{1 + \left( \frac{x}{\epsilon_H} \right)^2} - 1 \right)
	\end{align}
with $\epsilon_H = 0.001$, which is smooth and stable towards outliers like speckle noise.

\mymarginpar{regularizers}
The primary regularization term $R$ favors more plausible displacement fields by penalizing displacement change along
the scan pattern, leading to a more uniform displacement along the fast scan axis and limiting the change between B-scans. It is implemented as an additional cost term based on absolute displacement differences, where differences between B-scans are weighted lower to take the flyback time into account. Its influence is controlled by the weighting factor $\alpha$.
Since alignment is translation-invariant, 3 degrees of freedom are yet undetermined. They are fixed by the mean displacement term $M$, which penalizes the sum of displacements and thereby centers the alignment to the coordinate system's origin.
$T$ is only used in stage 1 of the optimization and normalizes tilt within B-scans. Since the regularization terms are not modified in this work, we refer the reader to the publications of Kraus et~al.~for details\cite{Kraus2012, Kraus2014}.

\mymarginpar{optimization, 2 pass \& multi resolution}
The objective function is minimized with a gradient-based L-BFGS solver. To avoid local minima, both optimizations are performed in a factor 2 multi resolution pyramid, beginning with the most downsampled volumes and zero-initialized displacements. The \textit{rough axial prealignment} only estimates axial displacements and tilt at B-scan granularity, whereas the \textit{3-D registration} estimates 3-D displacements for, in its last multi resolution level, all A-scans. The resulting displacements are then used for Hermitian resampling to obtain the, in terms of \figurename~\ref{fig:algorithm-overview}, \textit{prealigned} or \textit{motion corrected} volumes, respectively.

\mymarginpar{merging}
In the final \textit{merging} step, the motion corrected volumes are fused into a single \textit{merged volume} by a weighted average. Voxels with a sampling density higher than normal are assigned exponentially decaying weights. Weights below a small threshold are clipped to zero, while the remaining weights are normalized to achieve uniform brightness.
The sampling density-based weighting is done to achieve negligible weights for locations in the motion corrected volumes that correspond to areas not covered by the corresponding input volume due to transverse motion.
All motion corrected A-scans of such areas end up being sampled from one of the acquired input B-scans next to the gap, which would be sampled more frequently and replicated along the gap. At positions where gaps of both scans overlap, the replicated data of the B-scan with less oversampling remains, which typically originates from the scan with the smaller gap.

\subsection{Handling of saccadic motion artifacts in angiography data} \label{saccade_handling}

\mymarginpar{white lines}
OCTA data cannot be used for registration at locations where any of the two co-registered scans are affected by severe saccadic motion. Such motion causes the angiography signal to saturate, resulting in uniformly high intensities across the whole B-scan, as can be seen in \figurename~\ref{fig:white-line-removal}a, b. This means that anatomical features vanish and registration would be misguided. Instead, such B-scans are detected and excluded by a weighting factor that becomes zero at these locations.

\mymarginpar{input validity}
For this purpose, 1-D angiography validity masks $\vec v^X, \vec v^Y$ were defined for each input volume with values corresponding to B-scans, being 1 (valid), if neither the corresponding, nor the previous and succeeding 2 B-scans had a mean decorrelation value of 3 standard deviations above the mean, and 0 (invalid) otherwise. To match the image data during the multi-scale registration, this signal is downsampled in the same way as the intensity signals.

\mymarginpar{motion corrected validity}
Validity $\tilde v_{ij}^V$ of the \textit{motion corrected} OCTA A-scan $(i,j)$ of scan $V \in \{ X, Y \}$ is based on the validity of the sampling location in the \textit{input volume}, which is smoothly interpolated according to
\begin{equation} \label{eq:validity-resampling}
	\tilde v_{ij}^X(\mat D) = I\left(\vec v^X, y_j + \Delta y_{ij}^X \right) \qquad
	\tilde v_{ij}^Y(\mat D) = I\left(\vec v^Y, x_i + \Delta x_{ij}^Y \right)
\end{equation}
respectively, where the interpolator $I$ again uses hermite interpolation and thus provides continuous first order derivatives when $\tilde v_{ij}^V$ is used during optimization. Since the input validities only vary between B-scans, interpolation only depends on a single coordinate, the slow scan axis.

\subsection{En face OCTA image generation} \label{OCTA-projection}

\begin{figure}[!ht]
	\centering	\includegraphics{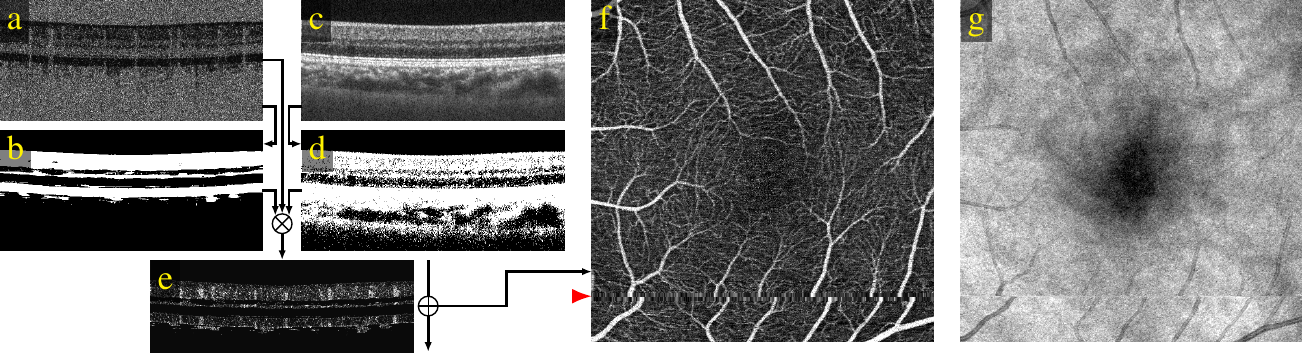}	\caption[Steps of the projection algorithm, healthy case]{		En face image generation in a 28 y/o healthy subject.
		(a) OCTA B-scan and thereof derived mask (b). (c) OCT B-scan and thereof derived amplitude decorrelation mask (d). (e) OCTA B-scan masked with both masks. After axial averaging, B-scans are arranged as en face image (f). (g) Corresponding en face OCT image.}	\label{fig:projection}\end{figure}

\mymarginpar{motivation \& input}
Because en face image generation is required to be fully automated, yet robust with respect to pathology, we opted against the typical approach of projecting between segmented retinal layer-boundaries.
Instead, our approach determines the voxels used for projection with a binary mask based on the contrast among neighboring OCTA intensities. The mask includes all voxels with relevant OCTA features in their neighborhood. The only local dependencies make the approach highly parallelizable and thus suitable for current processing hardware.
The projection algorithm is designed for amplitude decorrelation angiography data without removal of high decorrelation caused by background noise (section 4.1 in \cite{Jia2012}), which is deferred. In such volumes, low OCT signal areas like the vitreous or the deeper choroid in healthy subjects appear white, whereas low values occur only in the retina between vessels. \figurename~\ref{fig:projection} (a) shows a healthy example B-scan. The projection is performed on the axially prealigned volumes, see \figurename~\ref{fig:algorithm-overview}.

\begin{figure}[!ht]
	\centering	\includegraphics{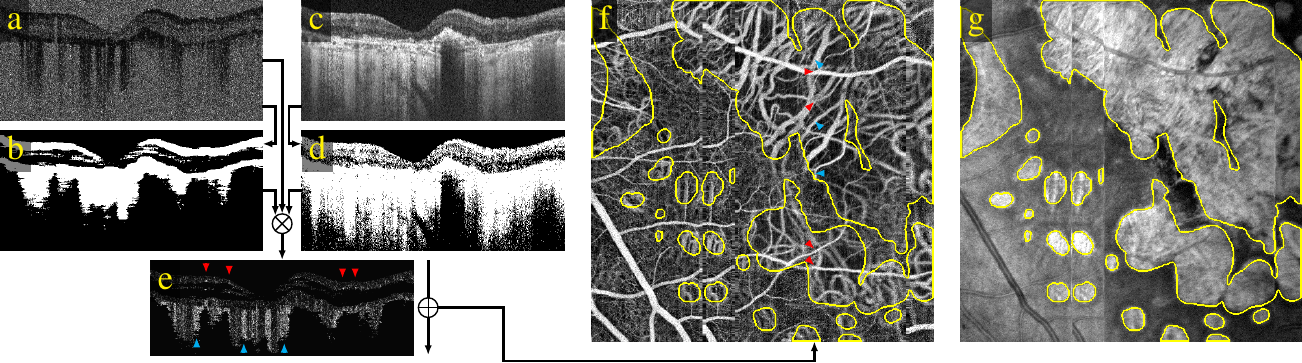}	\caption[Steps of the projection algorithm, pathologic case]{		En face image generation in an 83 y/o pathologic case with geographic atrophy (yellow).
		(a) OCTA B-scan and thereof derived mask (b). (c) OCT B-scan and typical thereof derived amplitude decorrelation mask (d). (e) OCTA B-scan masked with both masks. After axial averaging, B-scans are arranged as en face image (f), which shows choroidal vasculature in areas of atrophy. Corresponding superficial and choroidal vessels are marked with red and cyan triangles. (g) Corresponding en face OCT image.}
	\label{fig:projection-ga}
\end{figure}

\mymarginpar{steps}
After an initial 3-D median filter with radius 1 for noise reduction, 1-D rank-3 filters with width 15 are applied in the fast and slowly scanned directions subsequently. Through the choice of a low rank, these filters act similar to an erosion / minimum filter with added robustness toward outliers. The filter size is chosen to match the radius of large vessels, such that most of them are completely removed. The resulting volume typically has low values in the retina and high values elsewhere. After a normalization of the 5th and 95th percentile of the, in terms of $\vec v^X$ or $\vec v^Y$, valid B-scans to 0 and 1, thresholding is performed at 0.1 to form a binary voxel mask, see \figurename~\ref{fig:projection} (b). This mask essentially divides the volume in areas with and without relevant lateral contrast. Both this mask and the typical amplitude decorrelation mask derived via thresholding the OCT volume (d) are applied to the original OCTA B-scan (a), resulting in (e). Note that the OCT-based thresholding alone is insufficient, because it does not remove choroidal noise in the amplitude decorrelation angiography signal consistently which then manifests as varying brightness in the projection. Then, an en face image is formed by averaging the non-excluded voxels along depth,
and B-scan median subtraction is applied to compensate for increased OCTA signal due to bulk motion between decorrelated B-scans.
Finally, to prevent the use of saturated OCTA values during downsampling and interpolation, invalid B-scans according to $\vec v^X$ or $\vec v^Y$ were replaced with their nearest valid neighbor (red triangle), forming (f). Note that the contribution of these B-scans to the similarity term is mostly suppressed as detailed in the following sections. \figurename~\ref{fig:projection-ga} shows the same steps in an eye affected by age-related macular degeneration with geographic atrophy. In this case, due to increased light penetration in the area of atrophy, choroidal vasculature is additionally included in the projection, providing additional features for registration.

\mymarginpar{computational efficency}
In contrast to typical layer segmentation based approaches, this filtering-based method is highly parallelizable and thus suits GPU processing. Rank filters were implemented similarly to the median filter implementation of Perrot et~al.\cite{Perrot2014}, whereby the low rank filters disproportionally benefit from the therein described forgetful selection algorithm.
When using data from a specific system, the normalization mapping can be fixed to precomputed quantiles to reduce runtime.

\subsection{Incorporation of an angiography-based similarity measure} \label{OCTA-similarity}

\mymarginpar{new similarity term}
The new similarity measure is a combination of a structural $S_{\delta_0}^\text{str}$ and an angiographic $S_{\delta_0}^\text{ang}$ similarity term, which are defined according to
\begin{align} \label{eq:oct-similarity}
	S_{\delta_0}^\text{str}(\mat S^X, \mat S^Y, \mat D) &= \sum_i \sum_j \big(1 - \delta_{ij\delta_0}^{XY}(\mat D)\big) \cdot \sum_k L\left(R_{ijk}(\mat S^X, \mat S^Y, \mat D)\right) \\
	S_{\delta_0}^\text{ang}(\mat A^X, \mat A^Y, \mat D) &= \sum_i \sum_j \phantom{\big(1 -\ } \delta_{ij\delta_0}^{XY}(\mat D) \phantom{\big)} \cdot \ d \cdot L\left(R_{ijk}(\mat A^X, \mat A^Y, \mat D)\right), \label{eq:octa-similarity}
\end{align}
where $\mat A^X$, $\mat A^Y$ are the, according to \ref{OCTA-projection}, projected X- and Y-fast OCTA en face images, the loss $L$ is computed consistently with structural data by the pseudo huber norm as defined in Eq.~\eqref{pseudo-huber-norm} and
\begin{equation} \label{eq:validity-weight}
	\delta_{ij\delta_0}^{XY}(\mat D) = \delta_0 \cdot \tilde v_{ij}^X(\mat D) \cdot \tilde v_{ij}^Y(\mat D).
\end{equation}
is a weighting factor based on the OCTA validity of scans $X$ and $Y$ at the \textit{motion corrected} A-scan $(i,j)$. Thus, the similarity measures are identical except for the contrary weighting factors and that the summation along depth is replaced by a multiplication for the en face OCTA image. $\delta_{ij\delta_0}^{XY}(\mat D)$ becomes the default value $\delta_0$, if both angiography validities are valid; 0, if at least one is invalid and thus the difference based on saturated data is not meaningful for registration; and fades smoothly for in between sampling. Note that when $\delta_{ij\delta_0}^{XY}(\mat D)$ is zero, the angiography term becomes zero and the structural term becomes identical to the basic method. Due to regularization, these B-scans still benefit from the use of OCTA data in their neighborhood.

\mymarginpar{normalization $\eta$}
In their current form, the sum of both terms is still dependent on $\delta$, because the high contrast
in the en face OCTA image leads to higher residuals. Thus, when simply summing both terms, lower $\delta$ values, as they occur in saturated B-scans, would entail a lower loss, biasing the displacement towards sampling from such B-scans. To avoid this bias, the scaling factor $\eta$ is introduced to equalize both terms and the combined similarity term is defined as
\begin{align} \label{similarity-term-normalization}
	S_{\delta_0}(\mat S^X, \mat S^Y, \mat A^X, \mat A^Y, \mat D, \eta) = \eta \cdot S_{\delta_0}^\text{str}(\mat S^X, \mat S^Y, \mat D) + \frac1\eta \cdot S_{\delta_0}^\text{ang}(\mat A^X, \mat A^Y, \mat D).
\end{align}
The normalization is approximated by
\begin{align}
	\eta = \sqrt{S_{0.5}^\text{ang}(\mat A^X, \mat A^Y, \mat D_0) / S_{0.5}^\text{str}(\mat S^X, \mat S^Y, \mat D_0)},
\end{align}
where $\mat D_0$ is the initial displacement field of the current multi resolution level. Basing the normalization on the initial displacement field avoids recomputing the normalization in every iteration. This normalizes the structural and the angiographic data term to their geometric mean. With this choice, both data terms with their distinct sensitivity for misalignments equally influence the ratio between the overall similarity and the regularizers, thus reducing the dependency of the regularization strength on misalignment direction. Since both terms are normalized to the same value, the contrary weighting factors allow a bias-free exclusion of saturated OCTA data. $S_{\delta_0}(\mat S^X, \mat S^Y, \mat A^X, \mat A^Y, \mat D, \eta)$ replaces the OCT only similarity term in Eq.~\eqref{objective-function} in the \textit{3-D registration}, see \figurename~\ref{fig:algorithm-overview}. The en face OCTA data is not relevant for the \textit{rough axial prealignment}.

\subsection{OCTA volume merging} \label{octa-merging}

\mymarginpar{merging change}
While the merging approach for angiography data is based on the merging of OCT volumes described in \ref{oct-only-algorithm}, OCTA B-scans with saturated signal due to saccadic motion need to be excluded. Consistent with the en face OCTA image generation, the data of invalid B-scans according to $\vec v^X, \vec v^Y$ is first substituted by their nearest valid B-scan. This removes the influence of saturated B-scans during interpolation. The merging itself is extended by multiplying the averaging weights with $\tilde v_{ij}^V(\mat D)$ as previously described in section~\ref{saccade_handling} using the final displacements, to exclude the concerned OCTA B-scans from the merged result.

\begin{figure}[!ht]
	\centering	\includegraphics{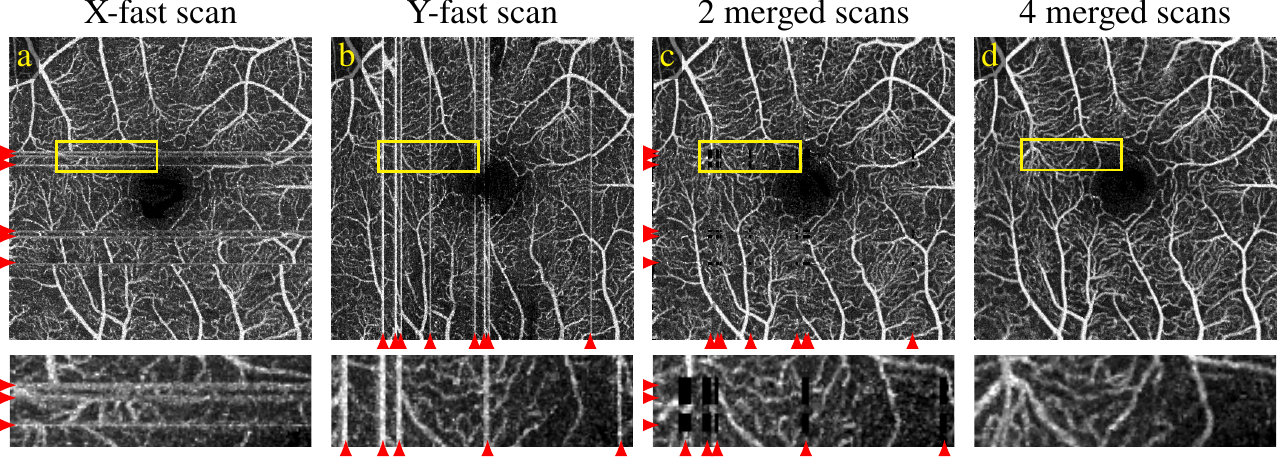}	\caption[Merged volume white line removal]{		Removal of white line artifacts in OCTA data that originate from saturated angiography signal due to saccadic eye motion. Shown are en face images of the superficial vascular plexus. Arrows point to white lines in the original scans, zoomed images show enlargements of the yellow rectangles.
		(a) Original X-fast and (b) Y-fast scan with white lines. Saccades are in-plane in the X-fast volume and thus partially corrected before the angiography computation.
		(c) Merged volume of one pair of orthogonal scans with white line B-scans excluded, data gaps appear as black boxes at the crossings of removed B-scans.
		(d) Merged volume of two pairs of orthogonal scans, no gaps remained.}	\label{fig:white-line-removal}	\end{figure}

\mymarginpar{visual description}
\figurename~\ref{fig:white-line-removal} visualizes the effect on en face projected volumes. Exclusion of white line B-scans removes the corresponding artifact. At these locations, only the valid data from the other scan remains. At the crossings of white lines from both scans, no valid data is available. While it would be possible to fall back to displaying saturated data, we chose not to do so to avoid confusion with vasculature, and display a small black gap instead. The issue can be avoided with high probability by registering additional scans in either or both scanning directions. Consequently, the degree of noise suppression depends on the actual number of merged volumes at the respective voxel.

\section{Evaluation framework}

\mymarginpar{eval overview}
Since ground truth motion is not available for real data, the motion correction error cannot be quantified directly. Instead, an evaluation framework is used, of which orthogonal evaluation aspects are described in each following subsection.

Section~\ref{objectives} introduces two objectives of the registration and their evaluation. Besides the \textit{alignment} between the co-registered scans, a \textit{reproducibility}-based schema to evaluate residual distortion is introduced.
Section~\ref{metrics} describes the separation of the error in its predominantly axial and transverse directional components in the quantitative analysis, by computing the \textit{ILM position disparity} and \textit{vessel disparity} metrics.
The OCT system and imaging protocol used for data acquisition and the study cohort, which contrasts \textit{elderly pathologic subjects} with \textit{young healthy controls}, is described in section~\ref{data}.
To conclude, section~\ref{evaluated-algorithms} gives remarks on the comparison of the algorithms \textit{OCT only} and \textit{OCT + OCTA}.

\subsection{Objectives of the registration} \label{objectives}

This section describes two objectives of registration based motion correction: the alignment between the co-registered scans and residual distortion within this alignment. For each objective, a schema is described that allows its evaluation by mapping the data of interest or thereof derived features to a co-registered space, allowing a pointwise comparison across the volume.

\subsubsection{Alignment} \label{alignment}

\mymarginpar{alignment objective}
With
alignment
we refer to the agreement of the two input scans after motion correction. This is a relevant objective because it is necessary to achieve a sharp merged image. Improper alignment will result in blurred or even doubled features, such as thin retinal layers or vessels.
\mymarginpar{alignment schema}
For evaluation, in terms of \figurename~\ref{fig:algorithm-overview}, \textit{input volumes} or thereof derived features are mapped to \textit{motion corrected} space by applying the estimated displacement fields, and then compared. For each eye and field size, alignment was evaluated for multiple acquired scan pairs.
While this kind of evaluation is important because it is closely related to the perceived image quality of the merged result, it does not quantify whether distortion was corrected and has the additional disadvantage that it is prone to favor overfitted registrations. Overfitting can happen when features that are only visible in one of the scans, like small capillaries or noise, are aligned with non-corresponding features in the other image. This can lead to a reduction of disparity metrics despite increased distortion.

\subsubsection{Reproducibility} \label{reproducibility}

\begin{figure}[!ht]
	\centering	\includegraphics{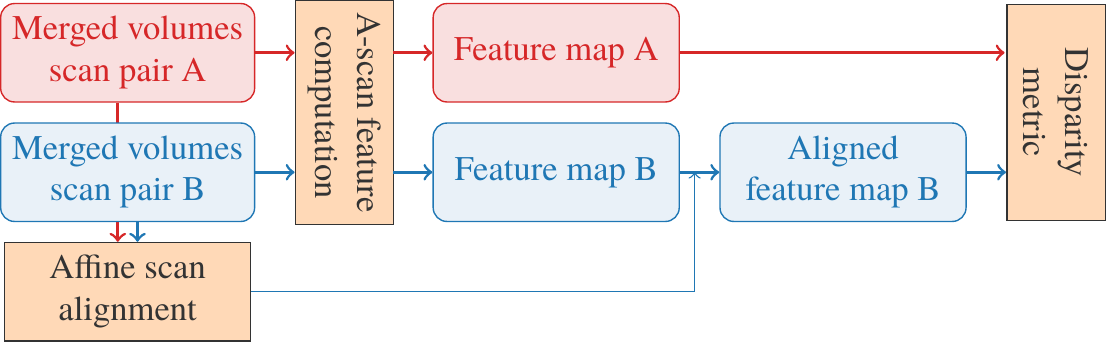}	\caption{Data flow of the quantitative reproducibility evaluation of independently acquired and independently motion-corrected volumes. After their computation, the A-scan features need to be aligned affinely before their disparity can be measured.}	\label{fig:dicorr-perf}\end{figure}

\mymarginpar{reproducibility objective}
The fundamental objective of motion correction is to minimize distortion. Repeated distortion-free OCT scans of the same eye should \textit{(re)produce} structurally consistent results up to scanner to pupil alignment, which thus qualifies as substitute objective.
\mymarginpar{reproducibility schema}
Based on these observations, the following evaluation schema was derived: Scan pairs of the same eye and scan protocol are acquired independently and are independently motion corrected and merged. Scanner alignment differences are estimated by registering each merged volume to the corresponding other volumes using a constrained affine registration. In addition to the translational and rotational components of rigid transforms, tilt-induced axial sheering along the transverse dimensions is compensated. Tilt occurs when the laser beam is not exactly aligned with the pupil \cite{Kraus2014}. This registration was performed consistently with the previous paper \cite{Kraus2014} on illumination corrected OCT data using a parameterization that resembles the aforementioned relevant degrees of freedom.
\figurename~\ref{fig:dicorr-perf} outlines the quantitative reproducibility analysis, which includes the feature map computations described in
section~\ref{metrics}.

Due to the rigid nature of the affine registration, there is no possibility for an overfitted co-registration and the independent eye motion during acquisition of each scan pair ensures that residual distortion differs. The noise reduction and artifact removal during merging does not bias the algorithm comparison because it was performed consistently, as described in section~\ref{evaluated-algorithms}.

\subsection{Feature-based, direction dependent disparity metrics} \label{metrics}

\mymarginpar{intensity = bad}
Various considerations were taken into account for choosing the metrics used during quantitative evaluation. To reduce the confounding influences of noise, which is rather high in OCT due to speckle, focus, and illumination inconsistencies, which appear at an increased prevalence in elderly and pathologic eyes, we opted for a comparison based on features of retinal structures
rather than an intensity-based one.
\mymarginpar{direction separation of errors}
Beyond that, certain features are only, or at least largely dependent on displacement along specific axes, allowing the evaluation to differentiate registration error components between these axes.
The two selected features, the algorithmically determined likeliness of an A-scan to show a vessel, and the position of the inner limiting membrane (ILM) along an A-scan, separate the error in its transverse and predominantly axial components. Note that evaluating these features only once per A-scan is sufficient to quantify the correction of motion-induced distortion, because this distortion only manifests between A-scans, not within them.
\mymarginpar{clinical relevance, reliability}
Further advantages of the selected features are their close relation to clinically relevant features and the high reliability in their automatic computation.
\mymarginpar{aggregation}
The final disparity score is computed by aggregating the per A-scan differences in the co-registered space defined by the respective evaluation schema in \ref{objectives}, usually via averaging.

\mymarginpar{layer segmentation theory}
Both features depend on layer boundaries, whose segmentation is described next. The chosen approach is similar to the method of Chiu et~al.\ \cite{Chiu2010}, which transforms each B-scan to a weighted graph with nodes corresponding to pixels and edges between neighboring nodes. The edge weights are computed from the axial intensity gradient, such that the shortest path from the beginning to the end of the B-scan corresponds to the boundary with the highest intensity difference between the surrounding layers. Our modified approach finds multiple boundaries at once to improve reliability by allowing the incorporation of relative layer depth constraints directly within the graph search. In detail, the graph construction is modified such that each node represents the pixels $z_1, ..., z_{n_\text{boundaries}}$ with $z_1 < ... < z_{n_\text{boundaries}}$ of the A-scan $x$, where pixel $(x,z_i)$ corresponds to the boundary position of the $i$-th segmented layer. Edges are added between nodes if the corresponding pixels of all boundaries are neighbors.
The edge weight is then determined by the sum of the edge weights of the individual layers. Knowledge about layer thickness was included by additionally constraining the minimum and maximum distance between consecutive boundaries within each node. Thus, a path along the nodes of the graph corresponds to $n_\text{boundaries}$ boundaries and each boundary is assigned with exactly one position per A-scan.

\begin{figure}[!ht]
	\centering	\includegraphics{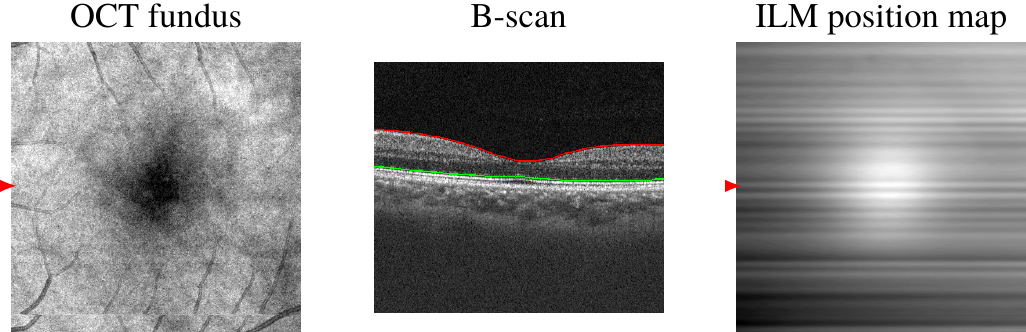}	\caption[Visualization of the layer segmentation]{Layer segmentation in a representative X-fast scan of a 28~y/o healthy subject. From left to right: OCT fundus image, B-scan at the position of the arrow in the fundus image with segmented ILM and EZ / IS-OS, ILM position map showing deeper ILM positions in white, as predominantly found in the fovea.}	\label{fig:layer-segmentation}\end{figure}

\mymarginpar{layer seg. impl.}
Based on this framework, layer segmentation was performed as follows. Initial ILM and posterior retinal pigment epithelium (RPE) boundary estimates are found by a joint segmentation. To decrease the computational demand of this initial estimate, the B-scans were blurred and subsampled by a factor of 5 in the axial dimension. Both layer estimates were subsequently smoothed with a Gaussian filter. Next, the volume was flattened with respect to the smooth RPE estimate
and a joint segmentation of
an estimate of the line immediately anterior to the ellipsoid zone (EZ) / inner segment-outer segment (IS-OS) junction
and the final posterior RPE boundary was performed in full resolution.
On the same flattened volume, the final EZ / IS-OS segmentation was performed in combination with the OS-RPE boundary. The ILM was segmented individually on a volume flattened according to the smooth ILM estimate. Finally, the layer positions were inversely shifted such that they again correspond to the true shape of the non-flattened volumes. A segmentation of the ILM and EZ / IS-OS is shown in a representative B-scan in \figurename~\ref{fig:layer-segmentation}.

\subsubsection{ILM position disparity}

An ILM position map is obtained by arranging the segmented ILM positions in all A-scans in an en face image as shown in \figurename~\ref{fig:layer-segmentation}. It is used to evaluate the predominantly axial registration accuracy. For this purpose, the ILM position maps of both compared scans are mapped to registered space before computing their A-scanwise difference and scaling it with the axial pixel spacing. The error metric thus describes the distance between the ILM segmentations in \textmu m.

\subsubsection{Vessel disparity}

A vessel map was used to quantify transverse registration performance, which was derived from a layer segmention-based en face OCTA image.
\mymarginpar{projection}
The en face image was computed by filtering the B-scans with a $3 \times 3$ median filter, followed by a mean projection between the ILM and the EZ / IS-OS (see \figurename~\ref{fig:layer-segmentation}).
EZ / IS-OS was selected for reliability of the segmentation in pathologies and because small errors in this avascular area have minimal effect on the projection.
\mymarginpar{white line removal}
Similar to the OCTA data similarity term, when alignment was evaluated, white lines caused by saccadic B-scans would degenerate the vessel probabilities. Thus, in accordance with section~\ref{saccade_handling}, such B-scans were detected based on the standard deviation of the original OCTA B-scans and replaced with their nearest valid neighbor in the en face image. This step is not performed when evaluating reproducibility, because in the therein compared merged volumes, such B-scans were already removed during volume merging (section~\ref{octa-merging}).
\mymarginpar{en face image normalization \& frangi}
Finally, again in both evaluations, the background noise was normalized by mapping the 10 and 95 percentiles to black and white, before computing the vesselness according to Frangi et~al.\ \cite{Frangi1998}. This filter is designed to emphasize bright tubular structures of multiple scales, corresponding to vessels of multiple diameters. Its result describes the likeliness of pixels (here: A-scans) to show a vessel, in a range from 0 to 1.
\begin{figure}[!ht]
	\centering	\includegraphics{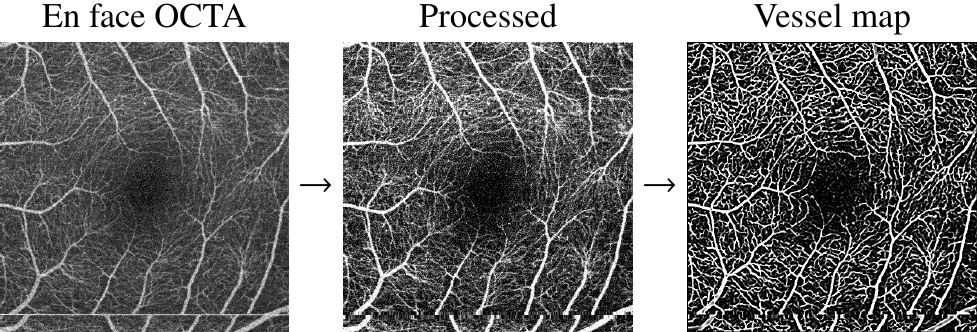}	\caption[Vessel map computation]{Vessel map computation in the same scan as in \figurename~\ref{fig:layer-segmentation}. From left to right: The en face OCTA image, the image after processing with saturated B-scans replaced, and the final vessel map.}	\label{fig:vessel-map}\end{figure}
\figurename~\ref{fig:vessel-map} shows the original mean projected en face OCTA image, the processed en face OCTA image after replacement of saccadic B-scans and normalization, and the resulting vessel map of a representative X-fast scan, as it is used in the alignment evaluation.
\mymarginpar{metric aggregation}
The error metric then describes the average difference in the co-registered vessel maps and is thus receptive to displacement-induced blurring and vessel doubling artifacts. The range of actually occurring scores is smaller than 1, because differences in perfect overlaps are still deteriorated by OCTA signal variation from e.g.\ pulse phase differences and, since individual vessels are not distinguished, misaligned vessels are not detected where they overlap with other vasculature.

\subsection{Imaging protocol \& study cohort} \label{data}

\mymarginpar{standard paragraph}
This study protocol was approved by the Institutional Review Boards at the Massachusetts Institute of Technology (MIT) and Tufts Medical Center. All participants were imaged in the ophthalmology clinic at the New England Eye Center (NEEC) and written informed consent was obtained prior to imaging. The research adhered to the Declaration of Helsinki and the Health Insurance Portability and Accountability Act.
\mymarginpar{system \& scan pattern description}
Data acquisition was performed with a research prototype developed at MIT, which was formerly described in \cite{Choi2013} and is briefly summarized in the following. The system uses a swept-source vertical cavity surface-emitting laser (VCSEL) light source with a 400 kHz A-scan rate at a central wavelength of 1050 nm. Raster scanning was performed in a grid of $500 \times 500$ A-scans with 5 repeats per B-scan, leading to a total scan time of \textasciitilde 3.9 seconds including flyback. Field sizes of both $3 \times 3$ \ts mm and $6 \times 6$ \ts mm were acquired for each eye, thus the lateral spacing is 6 \ts \textmu m and 12 \ts \textmu m respectively, whereas the axial spacing is 4.5~\textmu m in tissue. Repeated scans were registered rigidly with a cross-correlation based approach according to \cite{Fienup1990} at an accuracy of \textasciitilde 0.2 \ts \textmu m (1/32 pixel) before the OCTA signal was computed via full spectrum amplitude decorrelation angiography \cite{Jia2012}.

\begin{table}[!ht]
  \centering
 
  \begin{tabular}{r c l c c c} \hline
                      & \textbf{All}         &                                          &                  & \textbf{Alignment} & \textbf{Reproducib.}     \\
    \textbf{Category} & \textbf{\#cases}     & \textbf{Age}                             & \textbf{\#pairs} & \textbf{\#pairs}   & \textbf{\#merged pairs}  \\\hline
    NAION             &  1                   & 61                                       &   6              &  6                 &   6                      \\
    Dry AMD           &  1                   & 61                                       &   6              &  5                 &   5                      \\
    AMD with GA       &  1                   & 83                                       &   6              &  5                 &   6                      \\
    DM no DR          &  3                   & 62.0 \textpm{} 10.2 (52 -- 76)           &  18              & 17                 &  18                      \\
    PDR with DME      &  1                   & 67                                       &   6              &  6                 &   6                      \\

    \hline
    Pathologic        &  7                   & 65.5 \textpm{} 10.0 (52 -- 83)           &  42              & 39                 &  41                      \\
    Healthy           & 10                   & 27.1 \textpm{} \phantom{0}2.7 (23 -- 32) &  60              & 58                 &  59                      \\\hline
    All               & 17                   & 43.8 \textpm{} 19.9 (23 -- 83)           & 102              & 97                 & 100                      \\
    \hline
  \end{tabular}
  \caption[Pathology and age statistics of the study cohort]{Pathology and age statistics of the study cohort. Age is specified in mean $\pm$ standard deviation, with minimum and maximum in brackets.
 
  An
  additional 60~y/o NPDR case was excluded from evaluation because the retina left the scan area for wide areas in most scans, and it is thus not incorporated in the table.}
  \label{subjects}
\end{table}

\mymarginpar{imaging protocol \& subjects}
Three pairs of orthogonal scans centered at the fovea were scanned for each eye and field size. After each scan pair, the subject sat back and was realigned with the system to ensure that reproducibility is evaluated on independent acquisitions. If a subject blinked during an acquisition, that specific scan was repeated up to two times. In total, 18 eyes from 18 subjects were scanned. The study cohort, described in detail in \tablename~\ref{subjects}, included subjects with non-arteritic anterior ischemic optic neuropathy (NAION), dry age-related macular degeneration (AMD), AMD with geographic atrophy (GA), diabetes mellitus without diabetic retinopathy (DM no DR) and proliferative diabetic retinopathy (PDR) with diabetic macular edema (DME). An additional case with non-proliferative diabetic retinopathy (NPDR) was excluded from evaluation in its entirety due to severe axial motion that caused the retina to leave the scan over wide areas in most scans, which made further processing meaningless. It is thus not included in the table.

The pathologic cases are significantly older. Both factors make motion correction in this group harder due to opacities causing varying illumination and an increased number of blinks and saccades. Whereas the healthy subgroup had below 3 saccades per scan on average, it was more than 4 in the pathologic group, with several scans containing 10 or more.

\mymarginpar{excluded from quantitative}
A small fraction of scans needed to be excluded from the quantitative analysis due to failures of the layer segmentation.
In the dry AMD case, one scan contained a severe opacity that reduced the brightness of the superficial retina almost to background level, which caused the ILM segmentation to jump down to the RPE in this region instead. The scan pair thus needed to be removed from both alignment and reproducibility evaluation.
Similarly, the segmentation failed in two scans of a healthy subject due to an opacity. For one scan pair the segmentation error also occurred in the merged volume. Again, only the volumes with segmentation errors were removed from evaluation. However, in both subjects, the RPE was still visible, which seemed to suffice for registration purposes: visually there was no severe issue in axial registration in the excluded cases.
For a single scan each in the DM no DR and AMD with GA cases, despite the two rescans, no scan without a blink could be acquired. Consequently, the ILM could not be segmented properly in the corresponding areas. To avoid a negative bias, these scans needed to be excluded from alignment evaluation, but not from reproducibility evaluation, since the gaps were (automatically) filled by the other scan during merging of the therein used merged volumes.
In total, 97 of 102 scan pairs remained for alignment evaluation, and 100 of 102 merged scan pairs remained for reproducibility evaluation.
The higher yield in the merged volumes indicates that the reduced influence of opacities after merging and filling of blinks can improve the reliability of post processing steps like layer segmentation.

\subsection{Comparison of algorithms} \label{evaluated-algorithms}

The \textit{OCT only} method corresponds to the basic algorithm described in section~\ref{oct-only-algorithm}, whereas the \textit{OCT + OCTA} method includes the extensions proposed in this paper. Two adaptations were made to assure a fair and meaningful comparison. First, to separate the evaluation of registration accuracy from the effects of white line removal as described in section~\ref{octa-merging}, both the \textit{OCT only} and the \textit{OCT + OCTA} method use white line removal when merging OCTA data. Secondly, since the ratio between data term and regularization has a major impact on registration quality (see \figurename~\ref{fig:disparity_regularization}), the same scaling factor $\eta$ is also multiplied with the structural similarity term in the basic method analogously to Eq.~\eqref{similarity-term-normalization}. This ensures that identical regularization weights $\alpha$ correspond to equal ratios between data and regularization terms, and thus that the influence of this ratio is removed as a confounding variable when comparing the registration quality of both methods at identical $\alpha$ values.
Note that these adjustments make the basic algorithm equivalent to the proposed method with an angiography weight $\delta_0 = 0$. Statistical testing was performed with a two-sided Wilcoxon signed rank test at the 0.01 (**) and 0.001 (***) significance levels.

\section{Results}

We begin with a qualitative visualization of alignment in a case that is representative for the elderly and pathologic subgroup, where we demonstrate better co-registration, fewer artifacts and increased sharpness resulting from the approach with OCTA data. We proceed with a visualization of residual uncorrected distortion.
We then verify generalization of the observations quantitatively along the registration aspects highlighted in the previous chapter, including clinically relevant metrics that were also used in the relevant literature and an investigation of differences between the healthy and elderly pathologic subgroups.
We conclude the results with a peak into runtime.

\subsection{Qualitative results}

\begin{figure}[!ht]
	\centering	\includegraphics{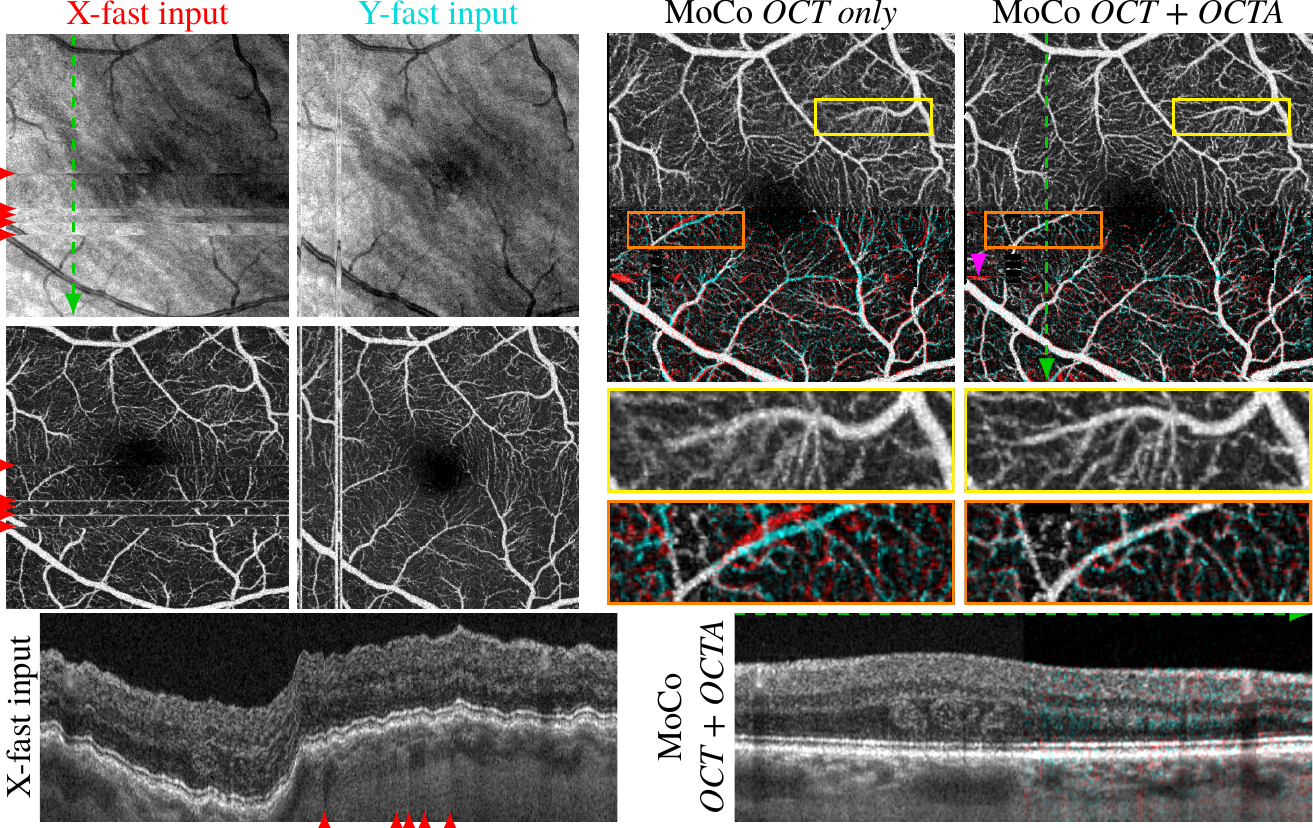}	\caption[Visualization of increased sharpness when using OCTA data during MoCo]{		Increased sharpness in merged data when using OCTA data during motion correction (MoCo) in the 61 y/o NAION subject. The merged images are split vertically and show the regular grayscale result in the upper half and a composite image in the lower half. Here, the contribution from the X-fast scan is displayed in red and that of the Y-fast scan in cyan. If the contributions from both scans are equal, these colors add up to grayscale. Areas with contribution from only one volume are also colored gray to not distract from misalignments. The composite images are median filtered to reduce distracting color from speckle noise and contrast enhanced to strengthen the color differences. The OCT B-scans are extracted along the dashed green arrows.}	\label{fig:qualitative_alignment}\end{figure}

\mymarginpar{alignment}
\figurename~\ref{fig:qualitative_alignment} demonstrates the potential of OCTA data in improving registration quality for an orthogonal scan pair by reducing or preventing artifacts that would otherwise compromise the merging benefits. Red triangles mark 5 saccades that happened in rapid succession during the X-fast scan.

\mymarginpar{transverse / en face}
The yellow box outlines an area of small misregistration in the \textit{OCT only} approach ($\alpha = 1$, justified below). Due to the misalignment, the larger horizontal vessel appears blurred, because, at the boundary, the vessel of one scan is merged with background in the other scan. Tiny structures like the vertical capillaries become hard to identify.
A more severe misregistration in the OCT only merged volume is outlined by the orange box, which originates from an area with relatively low contrast in the OCT input volumes. Since the misalignment exceeds the vessel diameter, the same vessel shows up twice, once from each merged scan. In the composite image, the coloring reveals the origin of both appearances, and that the red X-fast data is misaligned, which contains a saccade in close proximity.
The corresponding boxes in the \textit{OCT + OCTA} merged volume ($\alpha = 0.1$) do not show these artifacts. However, pointed out with the pink triangle, a small misregistration of the X-fast data between the two lowest saccades remains.

\mymarginpar{axial / B-scans}
The B-scan slices are extracted along the vertical axis (dashed arrows in en face images). Since the X-fast slice is oriented in slow scanning direction, it is distorted by severe axial motion. In the motion corrected slice, while no motion is expected in the cyan-colored Y-fast data, also the red X-fast data does not show axial motion artifacts, thus the overlay shows high agreement.

\mymarginpar{artifact frequency}
In the $3 \times 3$ \ts mm field size scans, results from 1 healthy and 7 pathologic eyes showed a double vessel artifact resulting from a single, small size misregistered area between saccadic motions. The distance between the double vessels is mostly dependent on saccade amplitude.
Furthermore, 3 pathologic results were heavily misregistered and not usable as a result of high amplitude saccades or a very high number of saccades ($\ge 12$ in both scans) combined with opacities. While occasional blurring occurred in the $6 \times 6$ \ts mm results as well, double vessels did not.

\begin{figure}[!ht]	\centering	\includegraphics{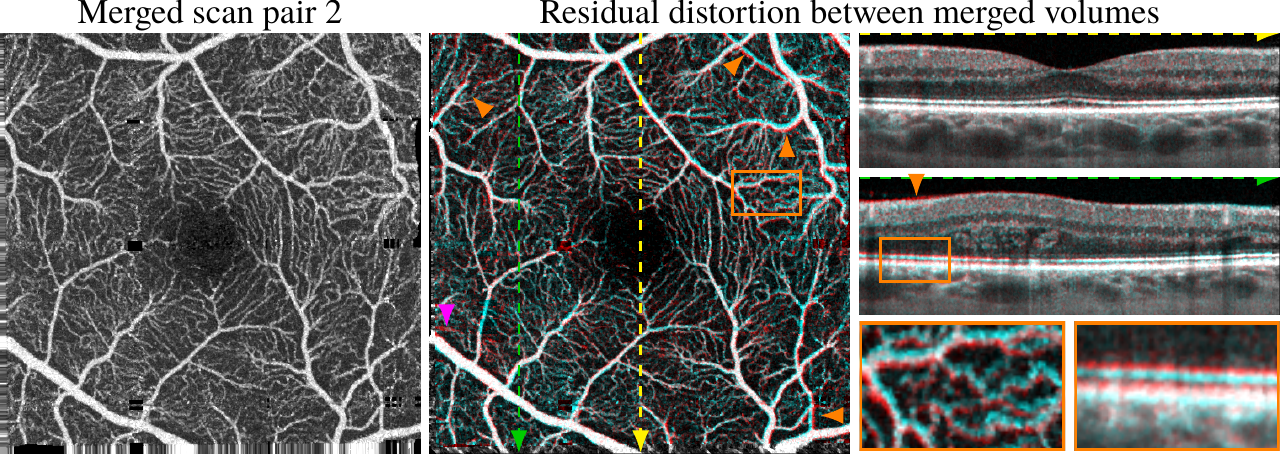}	\caption[Visualization of residual distortion]{		Residual distortion between independently acquired and independently
				\textit{OCT + OCTA}
		motion-corrected scan pairs of the same eye. The compared volumes are shown in red (merged volume
				in \figurename~\ref{fig:qualitative_alignment}) and cyan
		(left image).
				Zooms show distorted areas.
		}	\label{fig:qualitative_reproducibility}\end{figure}
\mymarginpar{reproducibility}
\figurename~\ref{fig:qualitative_reproducibility} visualizes the remaining distortion between the
\textit{OCT + OCTA} merged volume
in \figurename~\ref{fig:qualitative_alignment} and
an additional merged volume
from an independently scanned and motion corrected scan pair
(left),
which was, as a whole, affinely registered to the first as described in section~\ref{reproducibility}.
The
en face images of the merged volumes differ at the boundaries of certain vessels (orange markers). These differences reveal a 2 -- 3 pixel amplitude, low frequency residual transverse distortion manifesting in the peripheral areas. The pink triangle points to a location of misalignment in the first scan pair (see \figurename~\ref{fig:qualitative_alignment}), which also adds to the difference between the merged volumes. The B-scan extracted along the dashed green arrow shows the highest degree of axial distortion, which flips in direction roughly at the center of the scan and might be related to not fully corrected tilt differences caused by different pupil alignment.
Note that the affine registration achieves its best alignment in the center, where the extracted B-scan (dashed yellow arrow) shows excellent agreement. However, like affine registration of a single B-scan only, this is insufficient to capture the full volume's distortion, as apparent from the distortion in the other B-scan.

\subsection{Quantitative results}

\mymarginpar{overview}
We begin with plotting algorithm performance with respect to varying regularization weight $\alpha$ to justify our chosen algorithm configurations. We then continue with a boxplot subgroup analysis between young healthy and elderly pathologic cases.
In both figures, we show the evaluation objectives side by side, the metrics one below the other and plot both algorithms to give insights on how co-alignment vs. distortion correction (horizontally), axial vs. transverse error components (vertically) and the algorithms are influenced by the examined variables.

\begin{figure}[!ht]	\centering	\includegraphics[width=.8\linewidth,keepaspectratio]{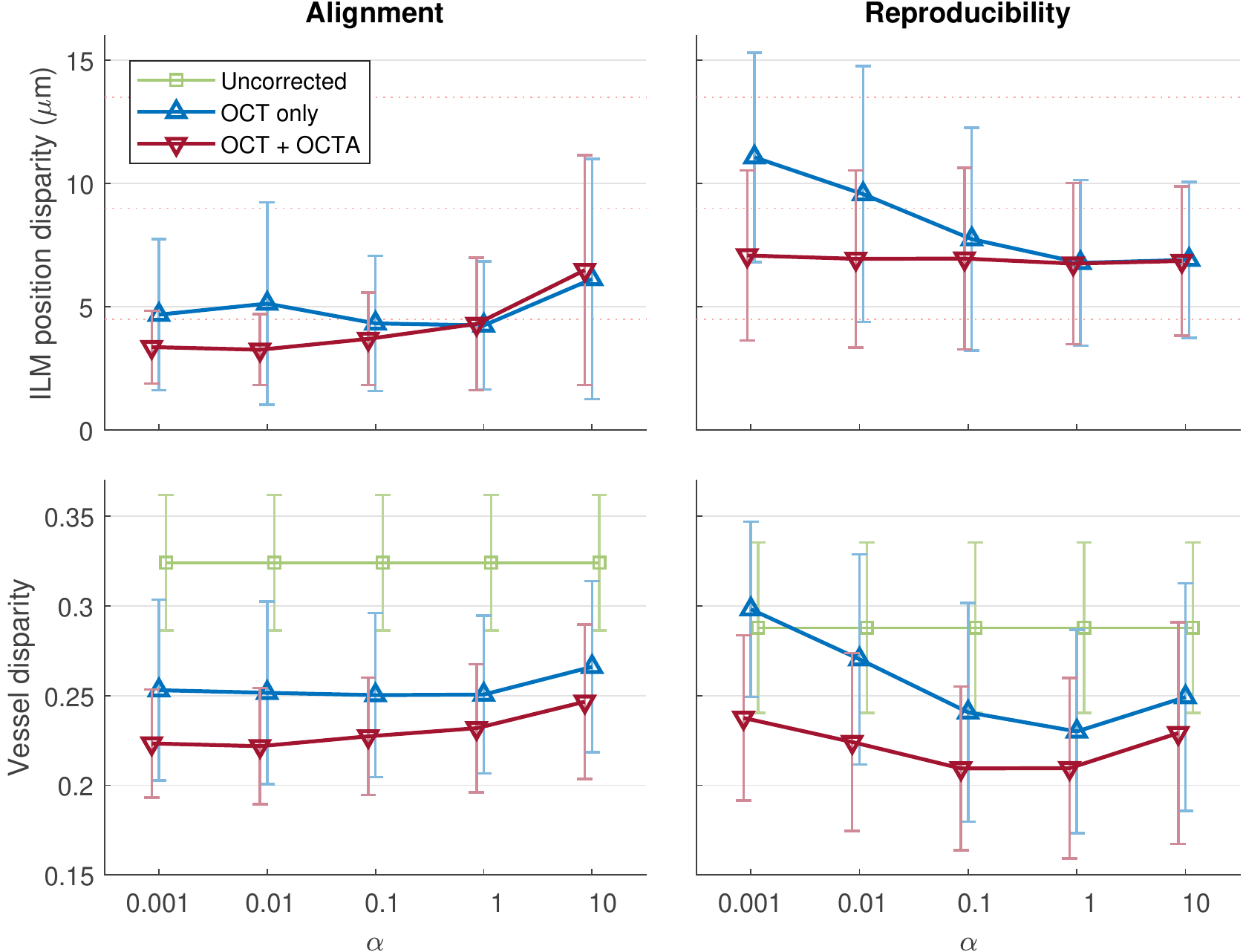}	\caption[Mean of mean feature disparity under varying degree of regularization]{		Mean of mean feature disparity under 5 degrees of regularization, specified by the weight $\alpha$ (slightly offset along the x-axis for visibility). Error bars indicate one standard deviation. For ILM position disparity, dotted grid lines show the 4.5~\textmu m pixel spacing. The uncorrected ILM position disparities, 140 and 45 \textmu m,
		are omitted for reasonable scaling. Uncorrected reproducibility was evaluated on the X-fast scans.}	\label{fig:disparity_regularization}\end{figure}

\mymarginpar{regularization weight}
\figurename~\ref{fig:disparity_regularization} shows the mean and standard deviation (over all scan pairs) of mean feature disparity (between the co-registered volumes) for motion correction with varying regularization weight $\alpha$. Both approaches clearly outperform the uncorrected data in terms of ILM position disparity. For reasonable configurations this also holds for vessel disparity, only for the smallest $\alpha$ value 0.001, the \textit{OCT only} approach shows slightly worse reproducibility. Furthermore, for equal regularization weights, the \textit{OCT + OCTA} motion correction consistently outperforms the \textit{OCT only} approach in terms of vessel disparity, and shows equivalent or lower ILM position disparity.
In terms of alignment, both measures show improvement with decreasing regularization weight. However, reproducibility results reveal that residual distortion increases again below a certain regularization weight, suggesting that improvements in alignment beyond this point stem from overfitting with implausible motion fields or wrong convergence.
Besides the general improvement in terms of vessel disparity, the \textit{OCT + OCTA} approach also seems to have improved stability, as mean vessel disparities are varying in a smaller range compared to \textit{OCT only} throughout the evaluated regularization weights, with lower standard deviations. The reason for the slightly lower best mean vessel disparities in reproducibility compared to alignment is that the herein compared merged volumes only have small residual gaps,
whereas the original scans used for alignment evaluation fall back to interpolated data for full white line B-scans (\figurename~\ref{fig:vessel-map}).

\mymarginpar{$\delta$ is not important}
In the \textit{OCT + OCTA} evaluations in \figurename~\ref{fig:disparity_regularization}, $\delta$ was fixed to 0.5. We also evaluated corresponding parameter setups with $\delta = 0.25$ and 0.75, but the difference was marginal (within 7\% standard deviation of $\delta = 0.5$). This is no surprise, because $\delta$ primarily affects the value of the objective function, but the location of the minimum, which corresponds to the optimal displacement field, remains where both the OCT and the OCTA data are aligned. Only for $\delta$ values close to 0, the contribution of the OCTA component becomes negligible compared to the OCT term's noise, and registration reduces to OCT only registration. The opposite happens for $\delta$ close to 1, with the additional drawback that, due to the lack of an axial dimension in the projected OCTA data, axial motion is not corrected anymore. Thus, $\delta = 0.5$ is the natural choice.

\begin{figure}[!ht]	\centering	\includegraphics[width=.8\linewidth,keepaspectratio]{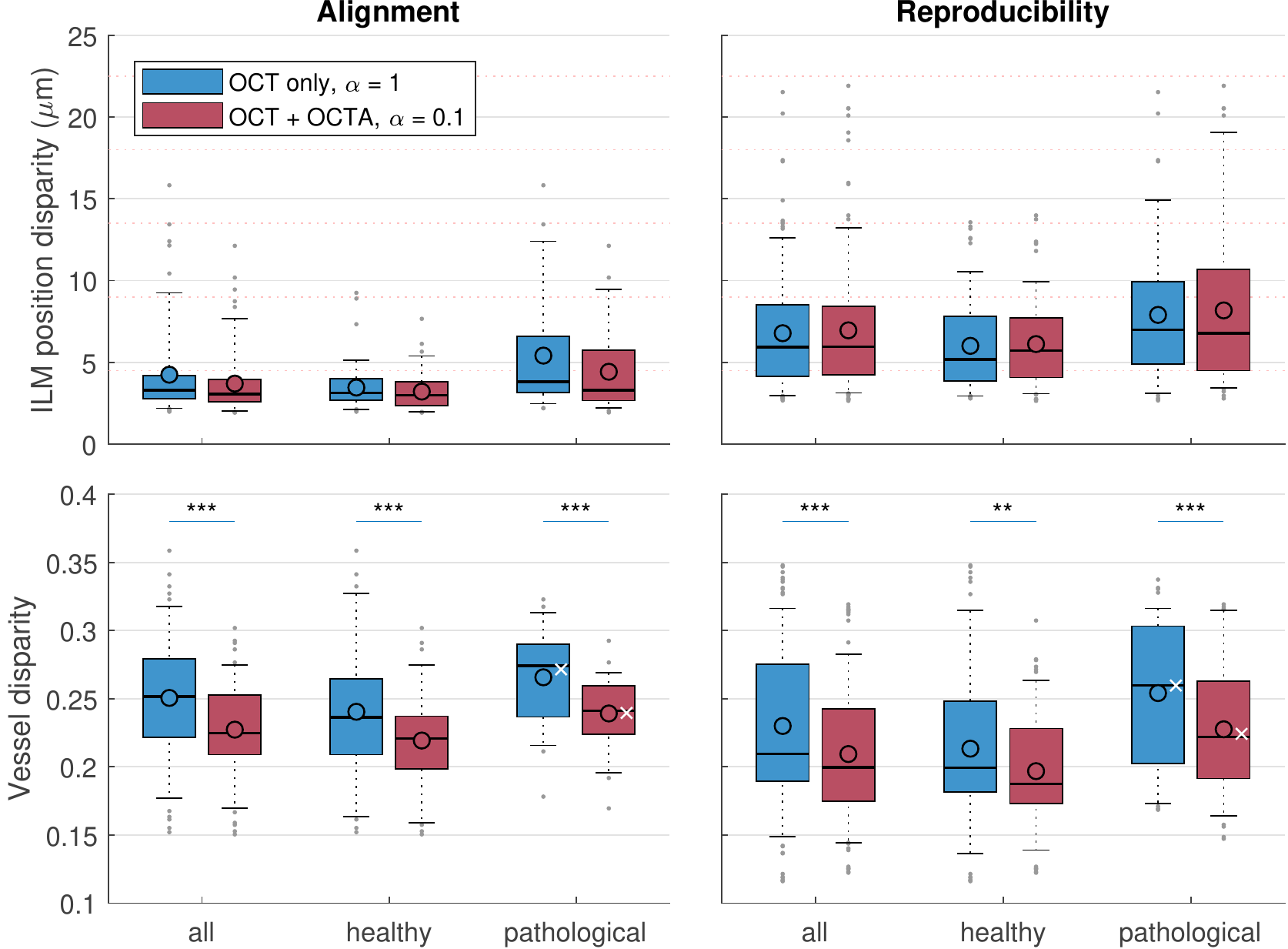}			\caption[Box plot of mean feature disparity in healthy and pathological cases]{		Box plot subgroup analysis of mean feature disparity. Whiskers represent the 5th and 95th percentiles. Circles show the mean of mean feature disparity analogously to \figurename~\ref{fig:disparity_regularization}. For ILM disparity, dotted grid lines show the 4.5~\textmu m pixel spacing. White crosses mark the scores of the scans shown in Figs.~\ref{fig:qualitative_alignment} and~\ref{fig:qualitative_reproducibility}.}	\label{fig:disparity_subgroups}\end{figure}

\mymarginpar{boxplot pathology}
\figurename~\ref{fig:disparity_subgroups} compares the healthy and the pathologic subgroups after motion correction with the best found configurations of the two algorithms. Since the $\alpha$ values 0.1 and 1 performed similarly in terms of reproducibility, we chose $\alpha = 0.1$ for the \textit{OCT + OCTA} approach for its better alignment scores. For the \textit{OCT only} approach, we decided for $\alpha = 1$ because the marginal improvement in vessel alignment using $\alpha = 0.1$ does not justify the loss of reproducibility.
Again, for ILM disparity, the results are very similar. In the alignment evaluation, only the pathological subgroup showed a notable improvement for the \textit{OCT + OCTA} approach. However, the improvement does not transfer to reproducibility and should thus be taken with a grain of salt. Vessel disparity shows a consistent improvement, which was found significant in both subgroups for both alignment and reproducibility. The boxplot also reveals that larger improvements in mean disparity are predominantly achieved in cases with comparatively worse results in the \textit{OCT only} approach: While the lower quartile and, for the healthy subgroup, the median show a small improvement when using OCTA data, the improvement is much more pronounced in the upper quartiles and the median of the pathological subgroup. Although using OCTA data reduced the performance gap between healthy and pathological cases, a significant difference remains. To connect the vessel disparity metric with a visual impression, we marked the scores of the scan pair shown in Figs.~\ref{fig:qualitative_alignment} and \ref{fig:qualitative_reproducibility} with white crosses.

\mymarginpar{something about numbers}
Tables~\ref{tab:alignment-perf} and \ref{tab:repro-perf} report metrics on alignment and reproducibility for the \textit{OCT + OCTA} algorithm, again using $\alpha = 0.1$ and $\delta = 0.5$.
Naturally, median errors are lower due to low influence of outliers, which is why we use the more comprehensive mean errors in our plots.
With 3.7 \ts \textmu m, mean ILM position disparity in terms of alignment using the \textit{OCT + OCTA} approach is below the pixel spacing (4.5~\textmu m). Metrics are higher for reproducibility (7.0 \ts \textmu m), and reveal that there is some residual distortion left on top of the misalignment.
\begin{table}[!ht]
  \centering
  \begin{tabular}{l ll c@{} ll c@{} ll} \hline
    \textbf{Alignment} & \multicolumn{2}{c}{\textbf{all}} && \multicolumn{2}{c}{\textbf{healthy}} && \multicolumn{2}{c}{\textbf{pathologic}} \\\cline{2-3}\cline{5-6}\cline{8-9}
   
    & uncorr.  & proposed      && uncorr.       & proposed      && uncorr.       & proposed      \\\hline
    median ILM & $135 \pm 106$ & $2.9 \pm 1.3$ && $126 \pm 101$ & $2.7 \pm 1.1$ && $148 \pm 111$ & $3.1 \pm 1.5$ \\
    mean ILM   & $140 \pm 104$ & $3.7 \pm 1.9$ && $128 \pm 101$ & $3.2 \pm 1.1$ && $157 \pm 109$ & $4.4 \pm 2.5$ \\
    mean vessel & $3.2 \pm 0.4$ & $2.3 \pm 0.3$ && $3.2 \pm 0.4$ & $2.2 \pm 0.3$ && $3.4 \pm 0.3$ & $2.4 \pm 0.3$ \\
    \hline
  \end{tabular}
  \caption[Disparity metrics on alignment]{Disparity metrics based on the X-fast and Y-fast volume alignment without (uncorrected) and with motion correction (proposed) specified as mean $\pm$ standard deviation over the volume-aggregated metrics. The metrics are median and mean ILM position disparity, both in \textmu m, and mean vessel disparity, which is scaled by 10.}
  \label{tab:alignment-perf}
\end{table}
\begin{table}[!ht]
  \centering
  \begin{tabular}{l ll c@{} ll c@{} ll} \hline
    \textbf{Reproducibility} & \multicolumn{2}{c}{\textbf{all}} && \multicolumn{2}{c}{\textbf{healthy}} && \multicolumn{2}{c}{\textbf{pathologic}} \\\cline{2-3}\cline{5-6}\cline{8-9}
   
    & uncorr.& proposed      && uncorr.     & proposed      && uncorr.     & proposed      \\\hline
    median ILM & $35 \pm 25$ & $5.8 \pm 3.3$ && $23 \pm 15$ & $5.2 \pm 2.3$ && $52 \pm 27$ & $6.7 \pm 4.1$ \\
    mean ILM   & $45 \pm 33$ & $7.0 \pm 3.7$ && $28 \pm 16$ & $6.1 \pm 2.5$ && $69 \pm 36$ & $8.2 \pm 4.7$ \\
    mean vessel & $2.9 \pm 0.5$ & $2.1 \pm 0.5$ && $2.7 \pm 0.5$ & $2.0 \pm 0.4$ && $3.1 \pm 0.3$ & $2.3 \pm 0.5$ \\
    \hline
  \end{tabular}
  \caption[Disparity metrics on reproducibility]{Disparity metrics based on the reproducibility of merged volumes (proposed) specified as mean $\pm$ standard deviation over the volume-aggregated metrics. Uncorrected results are based on the X-fast scans. The metrics are median and mean ILM position disparity, both in \textmu m, and mean vessel disparity, which is scaled by 10.}
  \label{tab:repro-perf}
\end{table}

\subsection{Runtime}

\mymarginpar{overall}
\tablename~\ref{tab:runtime} lists the runtime of the optimal configuration for varying volume sizes on a test system equipped with an Intel Core i7-6700K CPU, 32 \ts GB RAM and an Nvidia GeForce GTX 960 GPU (GM206, 4 \ts GB), which corresponds to current entry-level GPUs. The runtime statistic of the $500^2 \times 465$ voxel volumes is based on the full evaluation set when using the optimal configuration. Besides the GPU-implemented data similarity term evaluation and few exceptions during preprocessing, calculations are performed on the CPU and were not optimized. This includes the tilt estimation, which is only performed during prealignment. Thus, the relative scaling between the two registration steps is likely misleading. Furthermore, despite some memory copy bottlenecks were removed, the ongoing need for memory copies between CPU and GPU especially during optimization cause unnecessary overheads. In total, the GPU compute time makes up for only \textasciitilde 19\% of the runtime. There is thus potential for further speedup.

\begin{table}[h]
  \centering
  \begin{tabular}{l c l l l c l} \hline
  	\textbf{\#voxels} && \textbf{read / write} & \textbf{prealignment} & \textbf{3-D registration} && \textbf{total} \\\hline
    $500^2 \times 465$ && $11.5 \pm 0.9$ s & $26.7 \pm 0.2$ s & $31.4 \pm 8.0$ s && $86.1 \pm 8.0$ s \\
    $800^2 \times 433$ && $16.5 \pm 0.2$ s & $49.6 \pm 0.3$ s & $83.7 \pm 18.8$ s && $181.8 \pm 19.6$ s \\
    \hline
  \end{tabular}
  \caption[Runtime table]{Runtime of certain processing steps for different volume sizes. \textit{read / write} includes the time necessary to read the input from and write the merged volumes to disk, which is not relevant in a clinical setting. Registration timings include their respective preprocessing. \textit{Total} includes additional steps like saccade detection and merging.}
  \label{tab:runtime}
\end{table}

\mymarginpar{projection specifics}
While the 2-D en face OCTA data term evaluation is negligible compared to the volumetric OCT term, the runtime of the en face projection itself still must be shown to be small.
We stopped optimizing its GPU implementation at a runtime of \textasciitilde 209 \ts ms per volume, of which a large fraction is spent on sorting all voxels' intensities for the quantile-based normalization. Precomputing the normalization mapping for the specific system reduced the runtime to \textasciitilde 66 \ts ms.
We also ran the algorithm on 5 scan pairs with $800^2 \times 433$ voxels over a $12 \times 12$ \ts mm field-of-view to prove feasibility to even larger wide-field datasets with the same system. We used the same configuration up to a correspondingly larger displacement field ($800^2$ per volume) and a fixed normalization mapping for the OCTA projection.

\section{Discussion}

\mymarginpar{praise}
The results presented in this paper show that a sophisticated displacement model combined with the fine vascular features of OCTA can further improve post processing-based alignment and reproducibility at negligible additional computational demand. Overall scan distortion is reduced to micrometer scale and the achieved subpixel alignment improves visibility of capillaries and thin retinal layers.
\mymarginpar{improvement}
While both compared methods performed similarly well in the subgroup of younger healthy subjects, occasional blurred or misaligned vessels appeared predominantly in the elderly and pathologic group. Both these artifacts that stem from transverse misregistration were reduced in severity and frequency in the proposed \textit{OCT + OCTA} approach, which produces consistent, sharp images with higher reliability especially in the clinically more relevant subgroup.

\mymarginpar{better than the numbers}
Interpretation of the quantitative results must consider the whole evaluation pipeline, comprised of the acquisition with its underlying motion, the OCT scanner with its resolution and signal-to-noise ratio, layer segmentation and vessel map computation, the motion correction itself, the mapping of the feature maps and, in reproducibility evaluation, the affine registration. Thus, even perfect motion correction with a zero registration error cannot achieve zero disparity metrics.

\mymarginpar{importance of regularization changes}
The selection of the optimal regularization weight $\alpha$ should primarily be based on reproducibility, which evaluates the purpose of regularization: the validity of the displacement field.
Since ILM disparity did not vary significantly for the \textit{OCT + OCTA} approach (see \figurename~\ref{fig:disparity_regularization}), we chose $\alpha = 0.1$ based on vessel disparity as best configuration. However, since $\alpha = 1$ shows a similarly low disparity, the optimal choice is likely in between. In contrast, the \textit{OCT only} approach has its clear minimum at $\alpha = 1$. This suggests that when using OCTA data, the influence of the regularization can be lowered slightly. Potential explanations for this phenomenon are
the use of by-design brightness normalized amplitude decorrelation OCTA
and that the finer transverse features in OCTA data strengthen the importance of the data term in general. However, much more evident is that both approaches drastically lose reproducibility when reducing $\alpha$ below 0.1, stressing the importance of using appropriate regularization in OCT motion correction.

\mymarginpar{why pathologic are worse}
Although the gap between the young healthy and elderly pathologic groups could be reduced with the use of OCTA data, as shown in \figurename~\ref{fig:disparity_subgroups}, a significant difference between these groups remains. Because the proposed work uses a layer segmentation-free intensity-based similarity metric, we do not think that structural changes due to pathologies generally impede registration accuracy. While there are specific diseases that stop blood flow in larger regions and thereby cause a reduction of OCTA features, many pathologies introduce new structures, e.\,g. neovascularizations, or increase the visibility of choroidal vasculature, like geographic atrophy, which are automatically utilized for registration in our approach. Instead, we think that the gap is caused by a combination of the following reasons: First, the accuracy of the feature computations underlying the two disparity metrics could be negatively influenced by pathologies and the greater likeliness of vitreous opacities in elderly subjects. Secondly, since vitreous opacities cause reduced brightness at different regions in each scan, matching is complicated. And thirdly, an increased number of saccades not only increases the chance for misregistrations, but also makes the registration of the thus thinner saccade-free stripes harder. This is why some approaches reject them based on size, similarity with the co-registered scans or a combination thereof\cite{Hendargo2013, Ploner2019, Chen2017}.
Especially when considering the last two points, the results stress the importance of evaluating registration methods on elderly and pathologic data, which also has a higher clinical prevalence.

\mymarginpar{cmp Kraus14}
Kraus et~al.\ \cite{Kraus2014} used a similar evaluation pipeline. Besides different patient demographics, notable differences in their acquisition are a roughly half as long total scan time, significantly sparser transverse sampling and a smaller axial pixel spacing (3.1 vs 4.5~\textmu m). The latter is likely the main reason why they achieved a mean of mean ILM reproducibility error of slightly below 7 \textmu m, compared to the 7.1 \textmu m of the same method in this paper.
A further difference is that their blood vessel maps were extracted from the shadow artifacts in the inner segment-RPE region, which do not show capillary scale vasculature. En face OCTA images, like the one shown in \figurename~\ref{fig:qualitative_alignment}, show these fine structures, and thus necessitate higher transverse registration accuracy as highlighted by the yellow box. Combined with denser transverse sampling, the herein presented vessel disparity metric should be notably more responsive, but it is not directly comparable.

\mymarginpar{cmp Lezama16}
In contrast to the reproducibility-based evaluations, Lezama et~al.\ base their residual distortion evaluation on simulated distortion\cite{Lezama2016}. The comparison of individual scans to ground truth, the differently computed vessel metric, the motion simulation that is closely in line with the correction method and restricted to at most 4 saccades do not allow a direct comparison. In terms of alignment, a mean of median ILM disparity of 5.0 \ts \textmu m was reported in wide-field scans of healthy subjects, which is slightly above their pixel spacing. As common in OCT registration, their approach is based on an ILM segmentation, which bases the alignment on the few pixels that define a single layer transition. Our substantially lower disparity (2.7 \ts \textmu m in healthy subjects) might be a consequence of the use of intensity-based alignment as introduced in \cite{Kraus2012}, which depends on all pixels and thus all layer transitions along the A-scan, improving robustness towards noise and avoiding loss of spatial resolution by transverse smoothing. Despite they could not use OCTA data for registration, qualitative results show good transverse agreement with occasional discontinuities and double vessel artifacts in the composite OCT images. In the B-scan composite images of co-aligned scans, occasional bands of green and pink pixels above and below layer transitions like around the RPE reveal axial misalignment not present in our results (\figurename~\ref{fig:qualitative_alignment}).

\mymarginpar{cmp Chen17 (lissajous)}Chen et~al.\ evaluated their lissajous scan pattern based, OCT only, 3-D software motion correction approach qualitatively on healthy subjects\cite{Chen2017}. Besides good transverse alignment without double vessels, they achieved high reproducibility in en face OCT images.
While the distortion correction could be transferred to en face OCTA images in their follow-up paper on 2-D en face motion correction \cite{Chen2018}, smaller capillary-scale vasculature was degraded by blurring.

\mymarginpar{dual beam w/o good model is worse}
Note that the reproducibility of the representative pathological case in \figurename~\ref{fig:qualitative_reproducibility} is also much higher than the one shown by Kim et~al.\ \cite{Kim2020}, who used a scanner that acquires both orthogonal volumes at the same time. At the cost of increased system complexity, this simplifies the problem substantially, because only half the number of displacements need to be estimated.

Besides motion correction accuracy, clinical routine poses additional requirements on acquisition and compute time to reduce patient strain and to maintain patient throughput.
Both orthogonal and lissajous scans have a similar acquisition time due to oversampling by a factor of \textasciitilde 2 to enable registration.
However, the reported average compute times of similar volume size 3-D motion corrections, 70 \ts min (Chen et~al.) and 86 \ts s (ours) in dense regular field-of-view scans and 14 \ts min (Lezama et~al.) and 3.0 \ts min (ours) in wide-field scans vary substantially.

\mymarginpar{limitations}
Despite the demonstrated overall high accuracy and clinical applicability, the proposed method has limitations.
Naturally, compared to the methods of Kraus et~al.\ or Lezama et~al., our modified algorithm can only be used on OCTA datasets.
Furthermore, the current implementation is restricted to square volumes. Since non-square raster scans typically extend further in fast scan direction, the (square) overlapping part of two such orthogonal scans could be registered normally and the displacement field could be extended along the B-scans with minimal error.
Finally, a low amount of distortion can still remain and, as mentioned above, pathologic cases tend to a little worse motion correction results, which occasionally show up as blurring or double vessels.

\section{Conclusion}

We extended the orthogonal scan-based 3-D OCT motion correction method by Kraus et~al.\ to properly handle and utilize OCTA data.
We compared the proposed method quantitatively with the previous approach in a dataset with 102 scan pairs containing both healthy and a wide range of pathologic cases. Metric scores were connected with a visual impression by composite images of the most representative pathologic case.
We found a significant improvement in transverse alignment, which also transferred to significantly reduced residual distortion of the motion corrected results, especially in the clinically relevant elderly pathologic subgroup. Here, larger improvements were achieved, which demonstrate reliability of the presented method in pathology and bring the results closer to the control group despite more frequent and larger motion.
Furthermore, we showed that the additional steps have marginal impact on
runtime, preserving the clinical applicability of the algorithm.
The method has potential to further increase reproducibility of OCT(A)-derived disease metrics and to minimize motion induced inconsistencies between follow-up scans.

\section*{Funding}

\mymarginpar{need more of these}
Deutsche Forschungsgemeinschaft (MA 4898/12-1);
National Institutes of Health (5-R01-EY011289-31);
Retina Research Foundation;
Beckman-Argyros Award in Vision Research;
Champalimaud Vision Award;
Massachusetts Lions Eye Research Fund;
Macula Vision Research Foundation;
Research to Prevent Blindness.

\section*{Acknowledgements}
We thank Tobias Geimer, Ben Potsaid, ByungKun Lee, and Chen Lu for valuable discussions.

\section*{Disclosures}

SBP: IP related to VISTA-OCTA (P).
MFK: Optovue (C, P).
EMM: IP related to VISTA-OCTA (P).
NKW: Optovue (C), Carl Zeiss Meditec (F), Heidelberg (F), Nidek (F).
JSD: Optovue, Inc. (C, F), Topcon (C, F), Carl Zeiss Meditec (C, F).
JGF: Optovue (I, P), Topcon (F), IP related to VISTA-OCTA (P).

\bibliography{literature}

\pagebreak[4]
\section*{List of symbols}
\begin{itemize}
	\item $i,j,k$: horizontal, vertical \& axial voxel index \hfill see \ref{oct-only-algorithm}
	\item $w,h,d$: horizontal, vertical \& axial volume dimension \hfill see \ref{oct-only-algorithm}
	\item $X$: the X-fast scan (B-scans oriented horizontally) \hfill see \ref{oct-only-algorithm}
	\item $Y$: the Y-fast scan (B-scans oriented vertically) \hfill see \ref{oct-only-algorithm}
	\item $V$: an arbitrary scan \hfill see \ref{oct-only-algorithm}
	\item $\mat S^V$: the structural OCT input volume of scan $V$ \hfill see \ref{oct-only-algorithm}
	\item $\mat A^V$: the angiography / OCTA en face image, projected according to \ref{OCTA-projection}, of the input scan $V$ \hfill see \ref{OCTA-similarity}
	\item $\vec p_{ijk} = (x_i, y_j, z_k)$: voxel positions according to scan pattern \hfill see \ref{oct-only-algorithm}
	\item $\mat D$: displacement fields of all scans taking part in this motion correction \hfill see \ref{oct-only-algorithm}
	\item $\vec d_{ij}^V$: displacement vector at motion corrected A-scan $(i,j)$ \hfill see \ref{oct-only-algorithm}
	\item $\Delta x_{ij}^V, \Delta y_{ij}^V, \Delta z_{ij}^V$: displacement along x/y/z at motion corrected A-scan $(i,j)$ \hfill see \ref{oct-only-algorithm}
	\item $S_{\delta_0}$: the objective function's combined similarity term \hfill see \eqref{similarity-term-normalization} in \ref{OCTA-similarity}
	\item $S_{\delta_0}^\text{str}$: the objective function's structural similarity term \hfill see \eqref{objective-function} and \eqref{oct-only-similarity} in \ref{oct-only-algorithm} and \eqref{eq:oct-similarity} in \ref{OCTA-similarity}
	\item $S_{\delta_0}^\text{ang}$: the objective function's angiographic similarity term \hfill see \eqref{eq:octa-similarity} in \ref{OCTA-similarity}
	\item $R_{ijk}(\mat S^X, \mat S^Y, \mat D)$: the residual between the two input volumes after resampling to motion corrected volumes with displacement fields $\mat D$ \hfill see \eqref{eq:residual} in \ref{oct-only-algorithm}
	\item $L(\cdot)$: a loss function \hfill see \eqref{oct-only-similarity} in \ref{oct-only-algorithm}
	\item $L_{H,\epsilon_H}(\cdot)$: the pseudo-huber-loss from \cite{Kraus2014} \hfill see \eqref{pseudo-huber-norm} in \ref{oct-only-algorithm}
	\item $\epsilon_H$: parameter to control the smoothness of the pseudo-huber-loss \hfill see \eqref{pseudo-huber-norm} in \ref{oct-only-algorithm}
	\item $\vec v^V$: 1-D angiography B-scan validity mask of scan $V$ \hfill see \ref{saccade_handling}
	\item $\tilde v_{ij}^V(\mat D)$: validity mask of $V$ resampled to motion corrected A-scan $(i,j)$ \hfill see \eqref{eq:validity-resampling} in \ref{saccade_handling}
	\item $\delta_0 = 0.5$: default fractional weight of the angiography similarity term for a fully valid A-scan \hfill see \ref{OCTA-similarity}
	\item $\delta_{ij\delta_0}^{XY}(\mat D)$: fractional weight of the angiography similarity term at motion corrected A-scan $(i,j)$ during similarity term evaluation between scans $X$ and $Y$ \hfill see \eqref{eq:validity-weight} in \ref{OCTA-similarity}
	\item $\eta$: the scaling factor for normalization between the two similarity terms \hfill see \eqref{similarity-term-normalization} in \ref{OCTA-similarity}
	\item $R(\mat D)$: the objective function's displacement regularization term \hfill see \eqref{objective-function} in \ref{oct-only-algorithm}
	\item $\alpha$: the weight of the displacement regularization term \hfill see \eqref{objective-function} in \ref{oct-only-algorithm}
	\item $M(\mat D)$: the objective function's mean shift regularization term \hfill see \eqref{objective-function} in \ref{oct-only-algorithm}
	\item $T(\mat S^X, \mat S^Y, \mat D)$: the objective function's tilt normalization term \hfill see \eqref{objective-function} in \ref{oct-only-algorithm}
	\end{itemize}

\end{document}